\pdfoutput=1

\documentclass[%
    aps                ,   
    prl               ,   
    reprint            ,   
    twocolumn          ,   
    showpacs           ,   
    superscriptaddress ,   
    nofootinbib        ,   
    citeautoscript     ,   
    notitlepage        ,   
    floatfix
]{revtex4-1}
\usepackage[T1]{fontenc}
\usepackage[utf8]{inputenc}
\usepackage[english]{babel}
\setcitestyle{super}                 
\addto\captionsenglish{\renewcommand{\figurename}{Fig.}}
\makeatletter
\renewcommand*{\fnum@figure}{{\normalfont\bfseries \figurename~\thefigure}}
\renewcommand*{\@caption@fignum@sep}{\textbf{ | }}
\makeatother

\usepackage{amsmath}
\usepackage{amssymb}
\usepackage{braket}
\usepackage{accents}
\usepackage{siunitx}

\usepackage[hyperref]{xcolor}
\usepackage{graphicx}
\usepackage[export]{adjustbox}
\graphicspath{{figures/PNG/}{figures/PDF/}{figures/}}

\usepackage[
    unicode           = true  ,
    plainpages        = false , 
    pdfpagelabels     = true  , 
    bookmarks         = true  ,
    bookmarksnumbered = true  ,
    bookmarksopen     = true  ,
    breaklinks        = true  ,
    backref           = false ,
    colorlinks        = true  ,
    linkcolor         = blue  ,
    urlcolor          = blue  ,
    citecolor         = red   ,
    anchorcolor       = green ,
    hyperindex        = true  ,
    linktocpage       = true  ,
    hyperfigures      = true
]{hyperref}
\hypersetup{
    linkcolor         = [RGB]{000 114 189} ,  
    urlcolor          = [RGB]{000 114 189} ,  
    filecolor         = [RGB]{000 114 189} ,  
    citecolor         = [RGB]{217 083 025} ,  
    anchorcolor       = [RGB]{119 171 048} ,  
    pdftitle={Interpretable inverse-designed cavity for on-chip nonlinear and quantum optics},
    pdfauthor={Zhetao Jia <zhetao@berkeley.edu>}
}

\begin{document}

\title{Interpretable inverse-designed cavity for on-chip nonlinear and quantum optics}

\author{Zhetao Jia}
\affiliation{Department of Electrical Engineering and Computer Sciences, University of California at Berkeley, Berkeley, California 94720, USA}

\author{Wayesh Qarony}
\affiliation{Department of Electrical Engineering and Computer Sciences, University of California at Berkeley, Berkeley, California 94720, USA}

\author{Jagang Park}
\affiliation{Department of Electrical Engineering and Computer Sciences, University of California at Berkeley, Berkeley, California 94720, USA}

\author{Sean Hooten}
\affiliation{Hewlett Packard Labs, Hewlett Packard Enterprise, 820 N. McCarthy Blvd., Milpitas, California 95035, USA}

\author{Difan Wen}
\affiliation{Department of Electrical Engineering and Computer Sciences, University of California at Berkeley, Berkeley, California 94720, USA}
\affiliation{Applied Science and Technology Graduate Group, University of California at Berkeley, Berkeley, California 94720, USA}

\author{Yertay Zhiyenbayev}
\affiliation{Department of Electrical Engineering and Computer Sciences, University of California at Berkeley, Berkeley, California 94720, USA}

\author{Matteo Seclì}
\affiliation{Department of Electrical Engineering and Computer Sciences, University of California at Berkeley, Berkeley, California 94720, USA}

\author{Walid Redjem}
\affiliation{Department of Electrical Engineering and Computer Sciences, University of California at Berkeley, Berkeley, California 94720, USA}

\author{Scott Dhuey}
\affiliation{Molecular Foundry, Lawrence Berkeley National Laboratory, Berkeley, California 94720, USA}

\author{Adam Schwartzberg}
\affiliation{Molecular Foundry, Lawrence Berkeley National Laboratory, Berkeley, California 94720, USA}

\author{Eli Yablonovitch}
\affiliation{Department of Electrical Engineering and Computer Sciences, University of California at Berkeley, Berkeley, California 94720, USA}

\author{Boubacar Kanté}
 \email{Corresponding author: bkante@berkeley.edu}
\affiliation{Department of Electrical Engineering and Computer Sciences, University of California at Berkeley, Berkeley, California 94720, USA}
\affiliation{Materials Sciences Division, Lawrence Berkeley National Laboratory, Berkeley, California 94720, USA}

\date{\today}


\begin{abstract}
Inverse design is a powerful tool in wave-physics and in particular in photonics for compact, high-performance devices. To date, applications have mostly been limited to linear systems and it has rarely been investigated or demonstrated in the nonlinear regime. In addition, the "black box" nature of inverse design techniques has hindered the understanding of optimized inverse-designed structures. We propose an inverse design method with interpretable results to enhance the efficiency of on-chip photon generation rate through nonlinear processes by controlling the effective phase-matching conditions. We fabricate and characterize a compact, inverse-designed device using a silicon-on-insulator platform that allows a spontaneous four-wave mixing process to generate photon pairs at 1.1MHz with a coincidence to accidental ratio of 162. Our design method accounts for fabrication constraints and can be used for scalable quantum light sources in large-scale communication and computing applications.
\end{abstract}

\maketitle

\section{Introduction}

Enhancing nonlinear optical processes has been a long-standing challenge due to materials’ weak nonlinear response. The quest for effective approaches to achieve on-chip frequency conversion and generate photon pairs has been an enduring endeavor. Over the last decades, various nanophotonic platforms have been proposed to implement and enhance nonlinear photon generation processes, including wire waveguides \cite{fukuda2005,signorini2018}, nanobeam cavities \cite{lin2014}, metamaterials \cite{OBrien2015}, microring resonators\cite{Turner2008, Azzini2012, Engin2013, Kumar2014, Grassani2015, Preble2015, Ma2017, Mittal2018, Bruch2019, Guidry2020, Zhang2021, Ramesh2022}, periodically poled waveguides/cavities\cite{zhao2020a, lu2021}, and photonic crystal cavities\cite{marty2021, Kodigala2017, Contractor2022}. It is well-known that high-quality factor cavities designed at target frequencies can bolster the nonlinear process by enhancing the field with confinement. Yet, the effective phase matching conditions are typically challenging to satisfy in non-conventional cavity structures. Without effective phase-matching, the generated photons from different positions in nonlinear materials can destructively interfere, reducing the total generation efficiency. Such issues can be addressed by computational inverse design \cite{Borel2004, Burger2004, Preble2005, Molesky2018, Chakravarthi2020, Pestourie2022, Li2022, Jiang2021}. Recently, the adjoint method has been generalized to optimize nonlinear photonic processes, such as second harmonic generation or optical switches based on the Kerr effect \cite{lin2016, hughes2018, Khoram2019, Yang2023}. However, current optimization methods face difficulties in generalization to multi-photon generation processes, and the coupling efficiencies for both input and output channels are often overlooked due to the optimization complexity. In addition, an intuitive understanding of the inverse-designed structure is unclear. The optimized structure often lacks interpretability, making it challenging to gain insights into how and why the inverse design method works.\newline
To address these challenges, we put forth an inverse design approach to amplify the efficiency of on-chip photon pair generation. We implement this strategy using the open-source package \texttt{EmOpt} \cite{Michaels2018}. Our method employs a multi-frequency co-optimization strategy and calculates gradients with respect to the design parameters via the adjoint method. The resulting efficiency enhancement stems not only from the increased field intensity due to the confinement of light in high quality factor cavity resonances but also from the improvement of phase-matching conditions, along with coupling between the cavity and waveguide mode considered in the design. We demonstrate the capability of the proposed method by fabricating and characterizing an optimized device that enables the efficient generation of photon pairs. Interestingly, the shape of the proposed design can also be explained by an effective potential method, and the approximate solution aligns well with the finite-difference frequency domain (FDFD) simulation results. The proposed optimization strategy can be generalized to other nonlinear processes for compact frequency-mixing devices on-chip, and the performance can be further improved using global optimization methods \cite{Jiang2019, Hooten2021}.

\section{Method and proposed device}

\begin{figure*}
    \centering
    \includegraphics[width=0.9\textwidth]{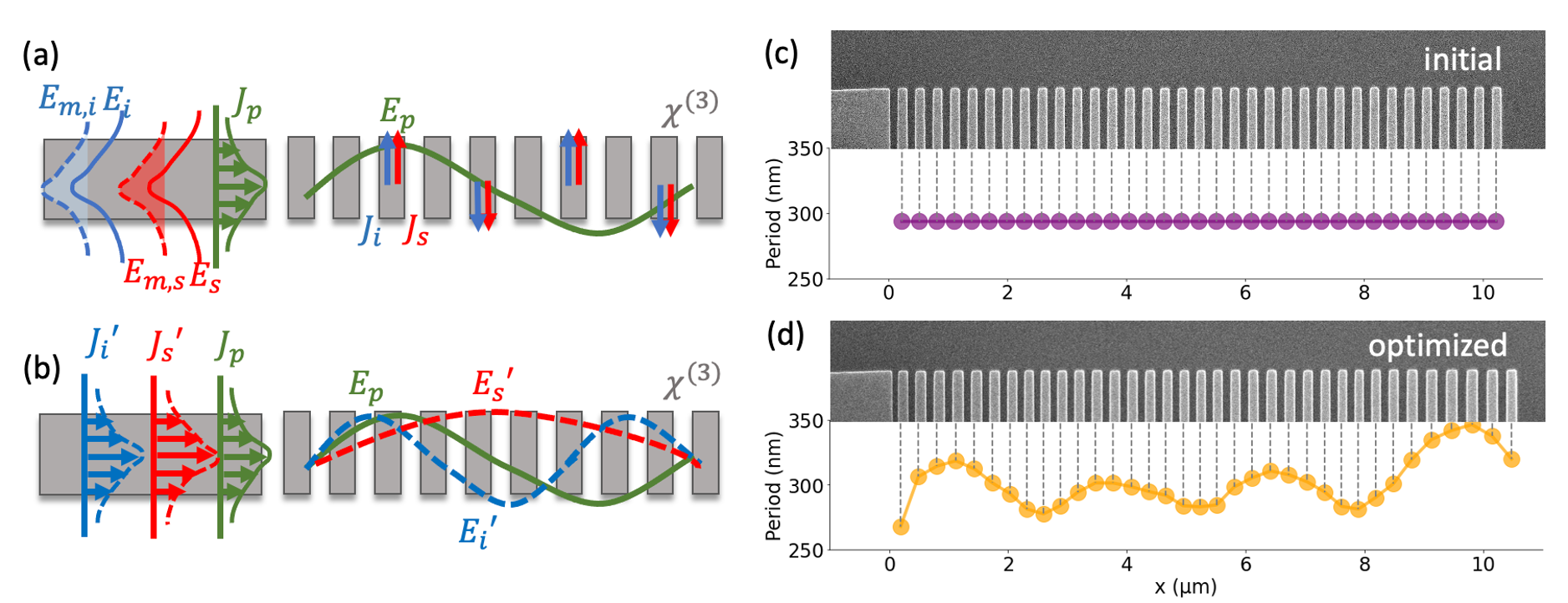}
    \caption{Multi-frequency co-optimization inverse-designed method and the fabricated device for an efficient spontaneous four-wave mixing process. (a) The pump ($J_p$) is injected as the fundamental mode of the waveguide, exciting dipoles of signal and idler frequencies ($J_s, J_i$) due to vacuum fluctuations. The radiation fields ($E_s, E_i$) of the dipoles are collected back into the waveguide. (b) In the adjoint process, pump, signal, and idler modes are injected into the waveguide and the phase matching of fields ($E_p, E_s', E_i'$) inside the cavity is calculated and optimized. (c-d) Scanning electron microscope image of the initial periodic grating structure before optimization (c) and apodized structure after optimization (d). The corresponding period of each grating is shown below the SEM image.
}
    \label{fig:fig1}
\end{figure*}

The proposed multi-frequency co-optimization method exemplifies the spontaneous four-wave mixing process shown in Fig.~\ref{fig:fig1}(a), where a single port is used to couple the pump/generated photons to/from the cavity. In the forward process, the fundamental mode at frequency $\omega_p$ is injected from the waveguide into the cavity, exciting the electric field distribution noted as $E_p$ in Fig.~\ref{fig:fig1}(a). Owing to the presence of the nonlinearity, vacuum fluctuations create dipole sources at other frequencies, specifically, signal and idler, shown as $J_s$ and $J_i$. The generated photons in the spontaneous four-wave mixing process are designed to be collected back to the same waveguide. The collection efficiency can be represented as a mode-matching integral between the collective radiation field generated by dipoles $E_s (E_i)$ and the waveguide mode at signal and idler frequencies $E_{m,s} (E_{m,i})$. Such nonlinear photon pair generation process can be approximated by the adjoint process shown in Fig.~\ref{fig:fig1}(b), where the adjoint sources $J_s', J_i'$, i.e., the fundamental mode of the waveguide at signal and idler frequencies $\omega_s$ and $\omega_i$, are reversely propagated back into the cavity. The efficiency can be represented in terms of the effective phase-matching integral as:

\begin{equation}
FOM = \left|\int_{cav} \chi^{(3)}(r) \beta(r) dr \right|^2,
\label{eq:sfwm_FOM}
\end{equation}
where $\beta(r) = E_p^{2} (\omega_p, r) E_s' (\omega_s, r) E_i'(\omega_i, r)$ is the effective phase-matching integrand, $\chi^{(3)}$ is the third-order nonlinear coefficient of the material, $E_p$ is the field distribution at pump frequency, while $E_s'$ and $E_i'$ are adjoint fields at signal and idler frequencies under continuous-wave (CW) excitation from the coupling waveguide. The integral is carried out over the cavity region. 

The proposed figure of merit can be interpreted as follows. First, the pump, signal, and idler frequencies ($\omega_p, \omega_s, \omega_i$) in Eq.~\eqref{eq:sfwm_FOM} can be selected in the CW-simulation to satisfy the energy conservation $\omega_s+\omega_i = 2\omega_p$. Second, the figure of merit uses the non-normalized electric fields ($E_p, E_s, E_i$), each obtained from a source excitation with fixed amplitude. The non-normalized field captures the cavity enhancement of the field intensities at the three frequencies, which also includes the coupling between the waveguide and the cavity for an efficient collection of the generated photons. In addition, with the overlap integral, the in-cavity phase-matching will be satisfied after the optimization to ensure the constructive interference of the generated photons.

\section{Optimization results}

\begin{figure*}[htbp]
    \centering
    \includegraphics[width=0.9\textwidth]{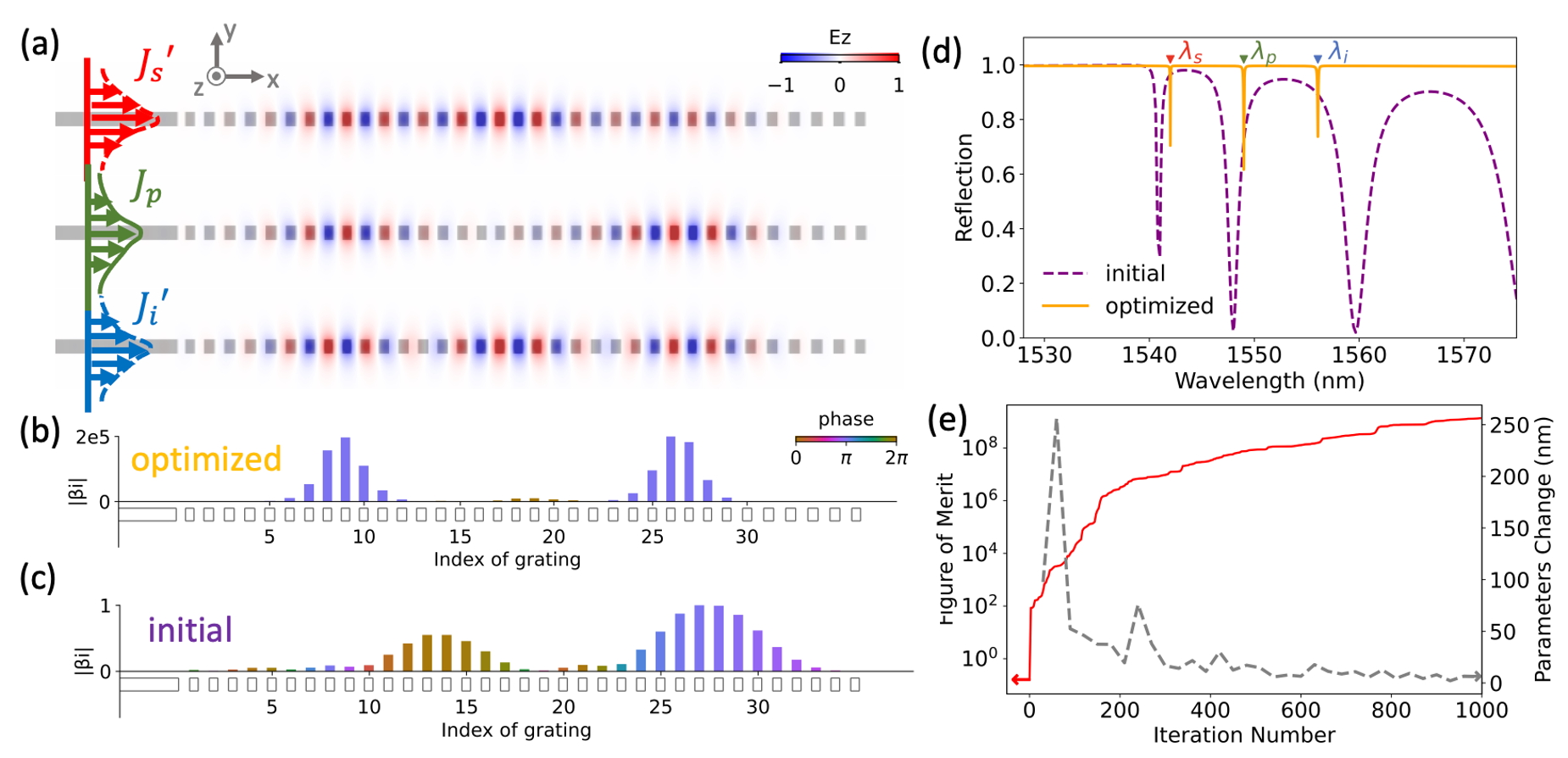}
    \caption{Optimization for nonlinear photonic structure. (a) Electric field profile ($E_z$) at the signal, pump, and idler frequencies for optimized structure, excited by fundamental modes ($J_s'$, $J_p$, $J_i'$) from the input waveguide.  (b-c) Illustration of the phase-matching condition. The height of the bar plot corresponds to normalized intensity, while the color indicates the phase of the phase-matching integrand, summed over each individual grating. The color consistency exemplifies the enhancement of phase-matching. (b) Optimized and (c) initial (periodic) grating. (d) Reflection spectra for grating before and after optimization. High-Q resonances, in alignment with target frequencies, are prominently observed. (e) The evolution of the figure of merit and parameters’ change over iterations during the optimization. 
    }
    \label{fig:fig2}
\end{figure*}

We adopt the hierarchical inverse design strategy, a two-step approach that proposes an initial physics-based guess, followed by a shape optimization using the adjoint method \cite{michaels2020}. Such strategy minimizes the computational cost by avoiding the large number of random guesses for initial conditions, and the fabrication limits can be easily enforced by adding simple shape constraints. Here, we perform the optimization using the width and gap of each grating, and the initial design is a periodic grating structure shown in Fig.~\ref{fig:fig1}(c) with the number of gratings $N = 35$, width and gap as $w = g = 147\mathrm{nm}$. The optimized grating structure is shown in Fig.~\ref{fig:fig1}(d) with three energy-matching resonant modes shown in Fig.~\ref{fig:fig2}(a). During the optimization, the fundamental TE mode is injected into the waveguide at the pump wavelength $\lambda_p = 1549 \mathrm{nm}$, and the field distribution $E_{p}$ is computed by the 2D FDFD solver of \texttt{EMopt}. The fundamental modes at signal and idler frequencies ($E_s', E_i'$) are excited by adjoint source ($J_s', J_i'$), also injected from the waveguide, at $\lambda_s =1542 \mathrm{nm}$ and $\lambda_i =1556 \mathrm{nm}$, respectively. The cavity enhancement of the fields can be visualized from the contrast of the field inside the cavity compared to that in the incident waveguide. The phase-matching integrand $\beta(r)$ in Eq. \eqref{eq:sfwm_FOM} is visualized as the bar plot in Fig.~\ref{fig:fig2} (b) and (c) for each grating. The heights of the bars stand for the amplitudes and the color for the phases. The field enhancement of five orders of magnitude is shown on the normalization of the y-axis after optimization. In the ideal case where all grating pitches are excited and contribute constructively to the photon generation, the phase (i.e., color) should be identical. Compared with the phase-matching plot for the initial period grating in Fig.~\ref{fig:fig2}(c), the phase-matching condition is greatly enhanced as the phase difference is minimized. The resonant frequencies of the modes can be probed as minima in the reflection spectrum, shown in Fig.~\ref{fig:fig2}(d). Initially, the resonances are not equally spaced, the quality factors are not high enough, and the phase-matching condition is not satisfied. After optimization, the three resonances are equally spaced in frequency, with sharper peaks in reflection due to the improved quality factors. The optimization uses the limited-memory BFGS (L-BFGS) algorithm \cite{Zhu1997} with hard constraints on the minimum width and gap to be larger than $110\mathrm{nm}$. The convergence of parameters can be observed from Fig.~\ref{fig:fig2}(e), with negligible parameters’ change after approximately 600 iterations, and the increment of the figure of merit is also shown.

\section{Measurement results}

\begin{figure*}[htbp]
    \centering
    \includegraphics[width=0.9\textwidth]{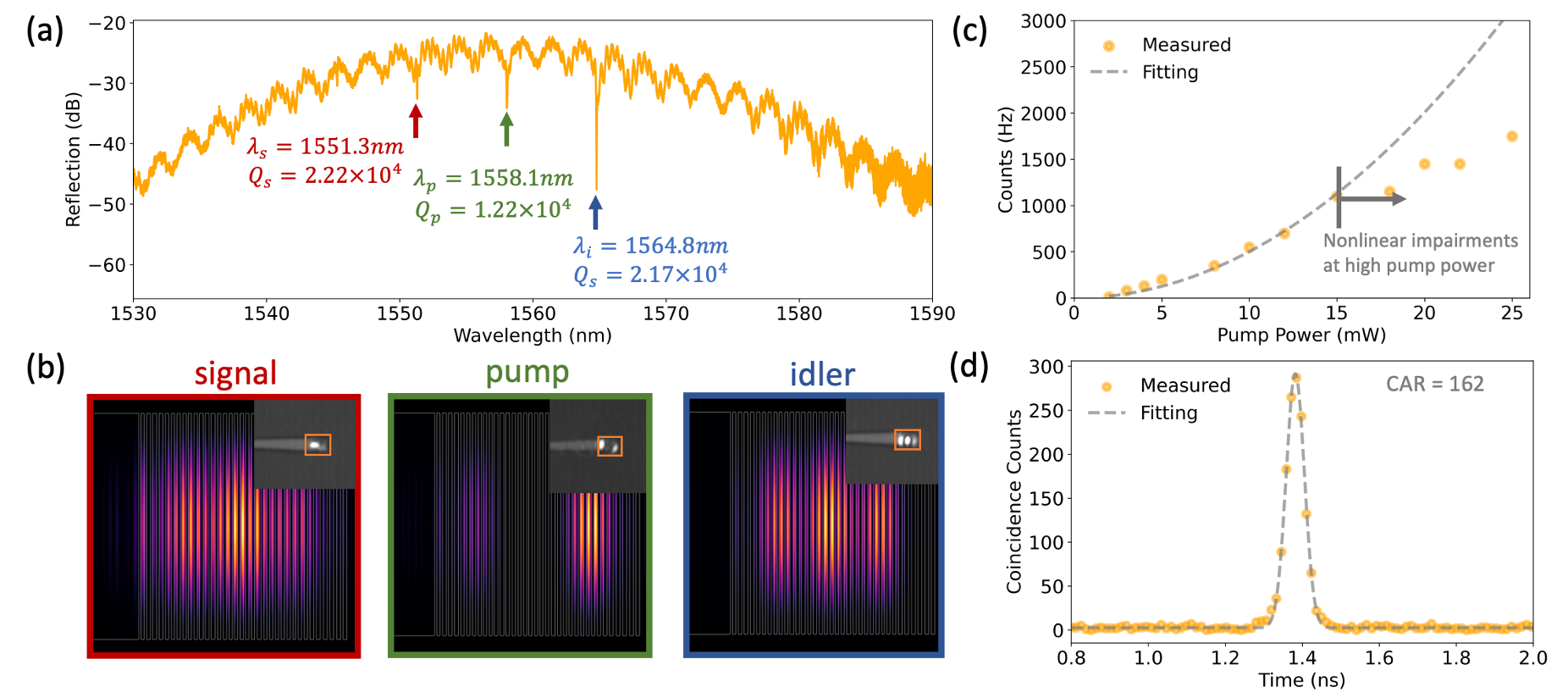}
    \caption{Measurement results of the inverse-designed device. (a) The reflection spectrum of the fabricated device clearly shows three distinct resonances for pump, signal, and idler frequencies, with their corresponding quality factors obtained by fitting. (b) Simulated scattered light captured by the objective lens monitored above the device and the observed camera images. (c) Light-light relationship for the spontaneous four-wave mixing process, with the collected data points aligning with the square-law fitting. The reduction of generation rate at high pump power is due to the appearance of other nonlinear effects. (d) Measured coincidence counts for signal and idler channels integrated over 10 minutes.}
    \label{fig:fig3}
\end{figure*}

As an experimental demonstration, we use conventional silicon on insulator (SOI) wafer with $220 \mu\mathrm{m}$ silicon layer with $3 \mu\mathrm{m}$ oxide box layer below. The device requires a single e-beam lithography and etching process followed by adding an oxide cladding of $1$um. We choose the transverse length of $10 \mu\mathrm{m}$ for simplicity in fabrication. The reflection spectrum of the inverse-designed device is measured by an optical spectrum analyzer in the linear regime, and the minima in reflection are extracted and fitted to obtain the loaded quality factor of pump, signal, and idler frequencies [Fig.~\ref{fig:fig3}(a)]. To confirm the field profile, the device is excited at resonance frequencies from a continuous wave laser source, and the scattered light is captured with an infrared camera as shown in Fig.~\ref{fig:fig3}(b). The obtained image is compared with the simulation, where the field is monitored $1 \mu\mathrm{m}$ above the device plane with Fourier components collected within the numerical aperture of the objective lens ($NA = 0.26$). The agreement between simulation and experiment confirms that the field distributions are optimized for phase-matching. Next, the nonlinear experiment characterizes the paired photon generation efficiency. The CW-laser is tuned to the pump wavelength, and the output power at signal frequency is monitored at different input power levels, shown in Fig.~\ref{fig:fig3}(c). The output power is proportional to the pump power squared, as expected for a degenerated spontaneous process before other nonlinear effects, such as free carrier or two-photon absorption, show up at around $15$mW \cite{Ma2020}. The quantum nature of photon-pair is confirmed by the intensity correlation measurement $g^{(2)}$ in Fig.~\ref{fig:fig3}(d), where the peak in the correlation indicates the photon pairs are generated simultaneously. The coincidence to accidental ratio is obtained from the fitted Gaussian curve as 162 and a maximum on-chip generation rate is $1.1\mathrm{MHz}$ at an on-chip pump power $0.78\mathrm{mW}$ after compensating for the loss. In the experiment, the signal and idler photons are filtered out with cascaded narrowband tunable filters, with $-120\mathrm{dB}$ extinction ratio for each channel.

\section{Interpretation of inverse-designed device}

\begin{figure*}[htbp]
    \centering
    \includegraphics[width=0.9\textwidth]{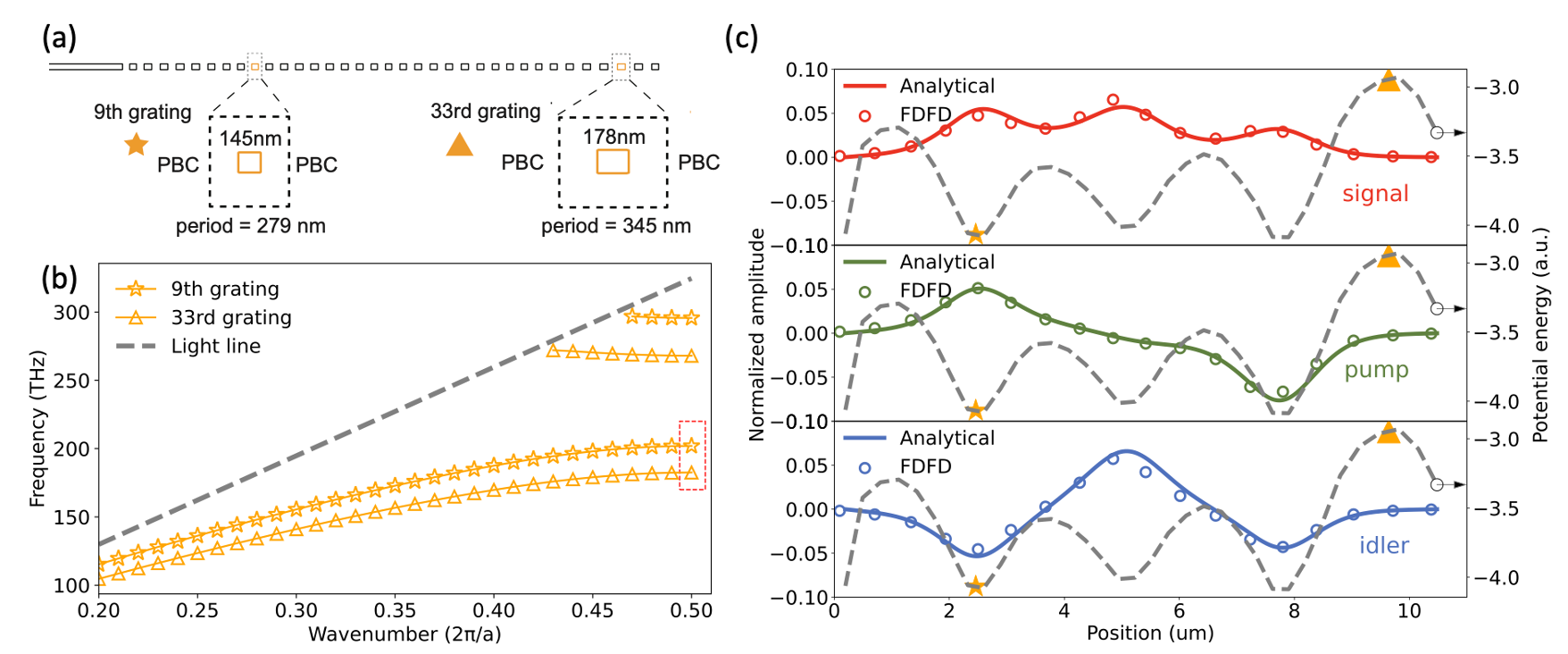}
    \caption{Interpretation of inverse designed cavity in the envelope function picture. (a) Cavity comprising perturbed gratings after optimization. Two example gratings with different periods are shown to calculate the band edge frequencies by applying periodic boundary conditions (PBCs) on left and right, and perfect matching layer (PML) at top and bottom. (b) Band diagram for the two selected gratings, where valence band edge frequencies are extracted. (c) The envelope functions (solid lines) of the three lowest modes in the effective potential landscape (dashed grey line) of the inverse-designed cavity. The amplitudes extracted from field profiles of FDFD simulation (circles) agree with the envelope function.}
    \label{fig:fig4}
\end{figure*}

The proposed cavity device can be seen as a quasi-one-dimensional photonic crystal with small perturbations. In that sense, our inverse design strategy can be understood as optimizing the perturbation and consequent mode field profiles to achieve maximal overlap integral while keeping the resonant frequencies equally spaced. Each mode field profile can be written as a product of the band edge mode $u(x)$ and a slowly varying envelope function $F(x)$; $E(x)=u(x)F(x)$. The envelope of a resonance mode $F(x)$ in a perturbed photonic crystal approximately follows the Wannier-like equation\cite{Charbonneau-Lefort2002,Painter2003},
\begin{equation}\label{eq:Wannier_main}
    -\left[\frac{1}{2m}\frac{\partial^2}{\partial x^2}+V_{\mathrm{eff}}(x)\right]F(x)=\omega^2 F(x)
\end{equation}
where $m$, $V_{\mathrm{eff}}$ and $\omega$ are the effective mass, effective potential, and the resonance frequency of the mode, respectively. The effective mass $m$ is defined by $m^{-1}\equiv \partial^2\omega_0/\partial k^2$ in analogy with that of electrons in solids, where $\omega_0$ is the photonic band frequency of the unperturbed photonic crystal. The effective local potential $V_{eff}$ can be extracted by simulating each grating with periodic boundary conditions (PBC). Two unit-cells with relatively small (star marker) and large (triangle marker) periods are highlighted as examples. The valence band edge frequencies are calculated using finite-element simulation, shown as the dashed box in Fig.~\ref{fig:fig4}(b). The calculation is performed over each of the gratings, and the obtained 1D effective potential is shown as dashed grey lines in Fig.~\ref{fig:fig4}(c), as the square of the valence band edge frequencies. The envelope function solutions for Eq.~\eqref{eq:Wannier_main} are plotted in Fig.~\ref{fig:fig4}(c). The envelopes show good agreement with the fields extracted from the FDFD solver in \texttt{EMopt}. Interestingly, the effective potential has three wells in the middle of the cavity region and a high wall near the right end. Therefore, the inverse-designed cavity can be interpreted as a three-coupled-resonator system between a highly reflective mirror on the right and an output coupler on the left. Compared with a single cavity case, the coupled-resonator configuration provides more degrees of freedom to adjust the optical mode shapes and, thus, a larger nonlinear overlap integral. In addition, our design method allows the device to have a smaller footprint for a given target resonance frequency, which is also advantageous for stronger light confinement.

\section{Discussion}
In summary, we propose an interpretable, computationally efficient optimization method for designing quantum and nonlinear photon generation devices on-chip. The method is demonstrated with a compact, fabrication-friendly, and highly reproducible device for photon pair generation in silicon photonics. The proposed method can be generalized for other nonlinear photon generation processes, for example, on-chip spontaneous parametric down-conversion, and third harmonic generation to cite a few. In addition, the target frequencies of the photon generation process can be chosen based on applications, which opens opportunities for inverse-design frequencies conversion across different bands, microwave-to-optic conversion, for example. The demonstrated photon-pair generation provides a feasible path for the compact integrated quantum light source. The design is also foundry compatible and can be used in scalable classical and quantum communication and computing applications based on integrated photonics platform.



\bibliography{references}


\end{document}